# 4D printing of mechanical metamaterials

Amir A. Zadpoor[1]

*Additive Manufacturing Laboratory, Department of Biomechanical Engineering, Delft University of Technology (TU Delft), Mekelweg 2, Delft 2628 CD, The Netherlands*

**ABSTRACT**

Mechanical metamaterials owe their extraordinary properties and functionalities to their micro-/nanoscale design of which shape, including both geometry and topology, is perhaps the most important aspect. 4D printing enables programmed, predictable, and precise change in the shape of mechanical metamaterials to achieve multi-functionality, adaptive properties, and the other types of desired behaviors that cannot be achieved using simple 3D printing. This paper presents an overview of 4D printing as applied to mechanical metamaterials. It starts by presenting a systematic definition of what 4D printing is and what shape aspects (*e.g.*, geometry, topology) are relevant for the 4D printing of mechanical metamaterials. Instead of focusing on different printing processes and materials, the paper addresses the most fundamental aspects of the shapeshifting behaviors required for transforming a flat construct to a target 3D shape (*i.e.*, 2D to 3D shapeshifting) or transforming a 3D shape to another 3D shape (*i.e.*, 3D to 3D shapeshifting). In either case, we will discuss the rigid-body shape morphing (*e.g.*, rigid origami) as well as deformable-body shapeshifting. The paper concludes with a discussion of the major challenges ahead of us for applying 4D printing to mechanical metamaterials and suggests several areas for future research.

---
[1] Corresponding author, email: a.a.zadpoor@tudelft.nl, tel: +31-15-2781021, fax: +31-15-2784717.

## 1. INTRODUCTION

Mechanical metamaterials [1] are architected materials that are designed to exhibit unusual properties and functionalities that are primarily resulting from the architecture of the material including the shape of their small-scale architecture [2-8] and the spatial distribution of different materials with differential properties [9, 10]. The list of the unusual properties and functionalities that can be achieved in this way has been rapidly growing during the last few years and includes not only negative [11-15] and extremal [16-19] properties but also adaptive behaviors [20-22], strain rate controlled properties [23], and device-like functionalities (*e.g.*, mechanical logic gates [24]).

Even though the rational design of the 'shape' of the small-scale architecture is only one of the many possible ways through which the unusual properties of mechanical metamaterials can be achieved, it has received by far the most attention. That is partially because the recent advances in 3D printing techniques have made it possible to manufacture architected materials with arbitrarily complex shapes. While other approaches, such as the spatial distribution of different materials, are also enabled by 3D printing, multi-material 3D printing approaches tend to be more expensive and limited in terms of the materials that can be processed and spatial resolutions that can be achieved. In addition to the ease of manufacturing, the rational design of metamaterial shapes is organically related to widely used computational approaches, such as topology optimization [25-31]. This could further streamline the design of the small-scale architecture of mechanical metamaterials and provide a sounds basis for property and functionality optimization of such materials. Because of both the abovementioned reasons, the rational design of 'shape' has emerged as the most widely used design paradigm in mechanical metamaterials.

Mechanical metamaterials whose properties and functionalities are dependent on the shape of their micro-architecture often need to be able to change their shape to achieve two important features, namely adaptive response [20-22, 32] and multi-functionality. It is clear why shape adaptation is required for a mechanical metamaterial to exhibit adaptive properties and functionalities. What may not be immediately clear is that multi-functionality may require shape adaptation too. For example, shape adaptation may be required because of the incompatibility between various fabrication techniques that are used to afford the architected metamaterial with multiple types of functionality (*e.g.*, see [33, 34]).

Regardless of whether shape adaptation is needed for creating adaptive properties or for achieving multi-functionality, it has to satisfy certain requirements. In particular, the change in the shape with time should be highly programmable, predictable, and precise. 4D printing is one of the very few options, if not the only realistic option, for the (large-scale) fabrication of shapeshifting mechanical metamaterials. In this paper, we will review 4D printing for application in the design and fabrication of mechanical metamaterials. Unlike previous reviews of 4D printing [35-38], which have been largely focused on materials and processes, this paper focuses on the fundamental aspects that are relevant for the 4D printing of mechanical metamaterials regardless of the type of the material and the specific additive manufacturing technology applied. In particular, we will focus on some conceptual insights, mathematical definitions and theorems, and specific examples of the application of 4D printing techniques in the context of mechanical metamaterials and machine-matter.

## 2. 4D PRINTING

Although the interest in creating shapeshifting objects is relatively recent, it precedes the recent advances in 3D/4D printing techniques [39-43]. The first attempts to fabricate shapeshifting objects used laborious manual approaches that, for example, required gluing multiple layers of polymers with mismatching thermal responses to create relatively simple shape transformations such as self-rolling or self-twisting. 4D printing [44-50], however, leverages the inherent advantages of 3D printing technologies to offer a fully automated process for fabricating functional objects with complex shape transformations. This could also enable the creation of the shapeshifting behavior at smaller (*i.e.*, micro/nano) length scales.

No clear, standard definition of 4D printing currently exists. I, therefore, propose the following definition to clarify the scope of this overview: "4D printing refers to single-step printing processes that result in objects whose shape (and function) could significantly change with time without significant change in their mass". This definition excludes the multi-step processes in which 3D printing is simply one of the several steps required for creating shapeshifting objects. The requirement to keep the mass (almost) unchanged excludes 3D printed objects whose changing shape is caused by (bio-)degradation. Moreover, as noted by other researchers [51-53], the shapeshifting process should result in a stable shape, which does not require external forces for its stability. In other words, the degree-of-freedom (DOF) of the 4D printed objects should be zero before and after shapeshifting [51, 53].

In practice, 4D printing is realized through the 3D printing of an object with programmed shapeshifting behavior. In other words, the object 'knows' how to change its shape. However, the shapeshifting behavior is not triggered until a stimulus is applied. The initiation stimulus could be light [54], heat [55-58], magnetic fields [59, 60], or the presence of water [61], among others. The stimuli differ in their working principle: while some stimuli provide (part of) the energy required for shapeshifting, some others only trigger the release of the energy that is already stored in the 4D printed object. For example, heat, light, and magnetic fields all provide some levels of energy for the shapeshifting action to take place. When applying these definitions and concepts in the context of mechanical metamaterials, some semantic questions arise. First, 'can mechanical forces be considered stimuli that activate shapeshifting in 4D printed objects?' Indeed, there is no fundamental difference between the direct application of mechanical force and the other stimuli that are used for the activation of 4D printed objects. The consequence of accepting the direct application of mechanical force as a legitimate stimulus for the activation of 4D printed objects is that it results in all 4D printed (soft) 'mechanisms' qualifying as 4D printed objects, which somewhat contradicts the usual understanding of what constitutes a 4D printed object. One could argue that the stimulus should work in a contactless (*i.e.*, untethered) manner. However, this additional requirement disqualifies stimuli such as pH or osmotic pressure that require direct contact with the 4D printed object to initiate the shapeshifting process. There is, therefore, limited fundamental ground to preclude the direct application of mechanical force as a legitimate stimulus. This point has some important implications for what we consider a 4D printed mechanical metamaterials, as many mechanical metamaterials are essentially 3D printed (soft) mechanisms.

Apart from the 3D printing process and material used in their fabrication, 4D printed objects are characterized by several variables including the initiation stimulus, the shapeshifting speed, the number of shapeshifting steps, and the reversibility/repeatability of the shape-shifting behavior. The type of stimulus that is required for the initiation of the shape shafting process is often strongly dependent on the type of material from which the object is printed. The speed and, thus, the times it takes for the shapeshifting to complete could be very different from a fraction of a second to several hours. Different applications impose different requirements regarding the shapeshifting time. For example, 4D printed (robotic) actuators often require faster response times as compared to drug delivery vehicles. Similarly, the reversibility and repeatability of the shape-shifting behavior are important for 4D printed actuators, while most other applications may not require that. Finally, complex changes in the shape may require sequential shapeshifting in which time delays are programmed into the 4D printed object such that it transforms its shape in multiple steps, thereby avoiding the collision of its different parts. Shape transformations achieved with 4D printing usually belong to one of the following categories: from one flat state (2D) to another flat state (2D/2D), from a flat state (2D) to a 3D shape (2D/3D), or from one 3D shape to another (3D/3D) (Figure 1). Transformations from a flat state to a 3D shape are sometimes referred to as 'self-folding origami' and have been studied more than both other types of shape transformations, perhaps due to the unique opportunities they offer. In particular, starting from a flat state facilitates the addition of surface-related functionalities [62] not only because surface (bio-)functionalization techniques usually work the best on flat surfaces but also because access to the entire surface may be only possible when starting from a flat state. Examples of surface-related functionalities that could be best applied in a flat state are nano-patterns that determine stem cell fate [63-65] and kill bacteria [66, 67], printed electronics that add sensors, actuators, and controllers to the 4D printed object [68], and superhydrophobic properties that could be used to create self-cleaning surfaces [69, 70].

## 3. SHAPE

Hidden in the definition of 4D printing is the implicit assumption that the shape of an objective is universally and objectively defined. A review of the literature shows that this may not be necessarily the case. In particular, there is a mismatch between how the shape is defined in the engineering and materials science communities on the one hand and the one used in the physics and mathematics community. Whereas in the former community the terms topology, geometry, and shape are sometimes used almost interchangeably and without very clear distinctions as to what constitutes each, the latter community adheres to very specific definitions of the concepts shape, geometry, and topology. The specific definitions used in mathematics have firm foundations in differential geometry and topology and are backed up by decades of research. We will, therefore, adopt those terminologies and will cover some of the relevant definitions and theorems from differential geometry and topology to put the concept of shape into the proper scientific context. In this approach, the shape is defined to have two important and related but distinct aspects, namely geometry and topology. The change in the shape of a 4D printed object might, therefore, mean a change in the geometry (Figure 1a), a change in the topology (Figure 1b), or a change in both geometry and topology of that object.

## 3.1. Geometry

Most shape aspects we know about an object relates to its geometry. Size, curvature, and angles all belong to the realm of geometry. Among the different geometrical aspects that could be considered, the curvature is perhaps the most well-defined concept and has received the most attention in the literature particularly in the context of how it relates to biology and bio-inspired design [71-78]. Moreover, many studies have shown that curvature is an important factor influencing the properties and behavior of biological [71-79] and materials science [80-83] systems. Finally, there are clear inherent relationships between both curvature and other geometrical concepts. For example, the Fenchel theorem [84] establishes an inherent coupling between size and the curvature of an object. Because of these reasons, in the remainder of this section, we will focus on curvature as a surrogate of the geometry of an object.

The curvature of a 3D surface is usually characterized using the concepts of Gaussian curvature and mean curvature. To calculate either one of those characteristics, we first need to find the principal curvatures of the surfaces, $\kappa_1$ and $\kappa_2$ (Figure 2a). The principal curvatures of a surface are the maximum and minimum values of the curvature of that surface and are determined within two orthogonal planes that pass through the point for which we are interested to calculate the curvature characteristics (Figure 2a). From this definition, it is clear that curvature is a local parameter and may (and usually does) change from one location within a 3D object to another location within the same object. Once the principal curvatures are determined, the Gaussian, $K$, and mean, $H$, curvatures are defined as:

$$K = \kappa_1 \, . \, \kappa_2, H = \frac{1}{2}(\kappa_1 + \kappa_2)$$

Depending on the value of the Gaussian curvature, the type of geometry that we are dealing with changes (Figure 2b). The geometry that we are most familiar with is Euclidean, where the Gaussian curvature is zero everywhere. A cylinder is a good example of an object belonging to the realm of Euclidean geometry. Since one of the principal curvatures is zero everywhere within a cylinder, the Gaussian curvature of a cylinder is always zero. When the Gaussian curvature is non-zero, the object is said to be intrinsically curved (at that particular point). Intrinsic curvature could be understood as the curvature that is sensed by individuals residing on that surface. For example, you know you are living on an intrinsically curved surface when the sum of the angles of a triangle is not $\pi$ anymore. If the value of the Gaussian curvature is positive at a particular point of a surface, the surface is said to be elliptic at that point. A sphere is elliptic everywhere, which is why elliptic geometry is sometimes called spherical geometry. On the other hand, a negative value of the Gaussian curvature implies the presence of a saddle-like shape, which belongs to the realm of hyperbolic geometry (Figure 2b).

The relationship between the different types of geometry is essential to understanding how 4D printing particularly 2D to 3D shape-shifting works. Everybody who has tried to wrap a piece of paper around a sphere knows that it is an impossible job. Similar to a saddle or any other surface with non-zero Gaussian curvature, a sphere is an example of a 'non-developable surface'. Whereas a developable surface can be folded from a flat object (without in-plane deformations being applied to that object), no surface with non-zero values of the Gaussian curvature can be folded from flat surfaces without distortions. Since many real-world objects of scientific and practical interest are non-developable, we need ways to get around this limitation. There are two major ways to bridge the Euclidean geometry with non-Euclidean geometries (*i.e.*, hyperbolic and elliptic). The first way is to induce in-plane deformation (*e.g.*, stretching) during the shapeshifting behavior of 4D printed objects. The other approach is to approximate the intrinsic curvature using an origami-inspired approach that mimics the curvature of non-developable using many rigid-origami folds. In this context, the qualifier 'rigid' refers to the fact that the paper is not stretched during the folding process and is only folded along its crease lines (*i.e.*, hinges). Several such examples are reviewed in [85]. This approach has been successfully applied to very complex hyperbolic objects, including triply periodic minimal surfaces (TPMS) [86]. While TPMS have attracted much scientific interest [87-91], they are very difficult to make from a flat surface. In TPMS, the mean surface curvature is zero everywhere. Mechanical metamaterials [92, 93] and meta-biomaterials [89, 94, 95] based on TPMS are, therefore, important examples of geometries where the mean curvature (in addition to the Gaussian curvature) is of critical importance.

## 3.2. Topology

Most aspects of what is usually considered to constitute the shape of an object are irrelevant when studying its topology. For example, changes in the dimensions, curvature distribution, or angles of an object do not change its topology. Indeed, topology is sometimes called "rubber-sheet geometry", because any shape transformation that only requires stretching and compressing the rubber sheet will not change its topology (tearing and gluing are not allowed). All such shapes are called 'homeomorphic'. For example, cylinders and spheres are homeomorphic to each other, while there exists no homeomorphic transformation to change the shape of a sphere to that of a torus. A sphere and a torus are, therefore, topologically distinct while there a sphere and a cylinder are topologically identical. An often-quoted joke says 'a topologist is someone who does not distinguish a donut from a coffee cup'. Tori with a different number of handles play an important role in defining the topology of objects. According to an important theorem from topology (*i.e.*, the classification theorem), each orientable surface is homeomorphic to a torus with a specific number of handles. In this context, orientability refers to the property of a surface that allows us to consistently define 'clockwise' on that surface and assign surface normal vectors to each point of the surface. Most surfaces of practical interest are orientable, meaning that most surfaces that we encounter in 4D printing are homeomorphic to a member of the torus family. The number of handles of a torus is called its 'genus', $g$, and is a topological invariant. Another topological invariant that is often used to describe the topology of an object is called the Euler characteristic, $\chi$. The genus and Euler characteristics are related to each other through the relationship $\chi = 2 - 2g$. The Euler characteristic of a sphere, which does not have any handles, is, therefore, 2 while the Euler characteristic of a single-handle torus is 0. The Euler characteristic of polyhedral can be calculated using the formula $\chi = V - E + F$ where $V$, $E$, and $F$ are the numbers of vertices, edges, and faces, respectively. It is intuitively clear that all convex polyhedra are homeomorphic to a sphere. It is, therefore, no accident that the Euler characteristic of all convex polyhedra is 2. This can be easily verified using the abovementioned relationship between the number of vertices, edges, and faces. This relationship is of great practical importance because shape triangulation similar to what is done in STL files allows us to approximate the Euler characteristic of any 3D object. As far as mechanical metamaterials are concerned, the topological properties of lattice structures are of particular interest given the fact that many metamaterials are designed using either beam-based or sheet-based lattice structures [8]. Examples of beam-based lattice structures include those based on polyhedral shapes while TPMS-based lattice structures are the prime examples of sheet-based lattice structures. The concept of the genus is very relevant for the study of TPMS lattice structures with different classes of minimal surfaces offering different values of the genus. As for beam-based lattice structures, the qualitative (*e.g.*, the pattern of the spatial distribution of node connectivity) and quantitative (*e.g.*, mean node connectivity number) nature of strut connectivity within the lattice structures are among the most important topological features.

### 3.3. Geometry-topology relationship

Even though geometry and topology describe two different aspects of the shape of an object, they are not fully independent of each other. The Gauss-Bonnet theorem is an important theorem from differential geometry that relates the sum of all Gaussian curvature within an object to its Euler characteristic. For physical objects of practical interest, the Gauss-Bonnet theorem can be simplified to the following relationship:

$$\int_O K dA = 2\pi . \chi(O)$$

where the integration is performed over the surface, $A$, of the object $O$. It then follows that for homeomorphic shapes with the same Euler characteristic, one could change the distribution of the curvature but not the sum of all curvature within that object. Figure 3 illustrates the consequences of the Gauss-Bonnet theorem for various objects with a genus of 0 (*i.e.*, zero-handle tori). Given that there are intrinsic couplings between topology and geometry, it is not possible to fully separate those two aspects from each other. Similar illustrations for objects with a higher value of the genus further clarify the relationship between the curvature distribution and genus (Figure 3b).

In terms of 4D printing technologies, the vast majority of the studies performed to date have been focused on changing the geometrical features of 4D printed objects. There has been almost no attempt to specifically address the shapeshifting in terms of the topological features of 3D objects. Of course, it is more challenging to change the fundamental shape features of mechanical metamaterials including the genus of a TPMS-based lattice structure or the node connectivity of a beam-based lattice structure, because that requires creating and/or removing functional connections between various places within the mechanical metamaterial.

### 4. 2D TO 3D SHAPE TRANSFORMATION

The most common type of 4D printing shapeshifting technologies pursued by researchers has been 2D to 3D shape-shifting where an initially flat object transforms its shape into a 3D object. There are several categories of practical applications for 2D to 3D shapeshifting behaviors. Most importantly, starting from a flat shape allows for integrating functionalities that are either surface-related (*e.g.*, surface nanopatterns [96-98], superhydrophobicity/superhydrophilicity [99-102]), thus requiring access to the entire surface area of the object, or have to be created using technologies that have traditionally only worked on flat surfaces (*e.g.*, nanolithography or electronic device fabrication). As far as mechanical metamaterials are considered, the incorporation of certain surface-related functionalities is crucial for creating multi-functional metamaterials that could be used for biomedical or certain high-tech industries. Surface nanopatterns could be used to guide the differentiation of stem cells [65, 103, 104], to kill bacteria [105-109], thereby preventing medical device-associated infections, or to induce hydrophobicity that can be used for creating self-cleaning surfaces [110]. Another important area where a flat starting point could be the key to the successful incorporation of additional functionalities is when the integration of electronic devices is needed. Mechanical metamaterials that are aimed for application in (soft) robotics, medical devices, and embedded systems often require the presence of electronic components, such as sensors, actuators, and control circuits. Given the fact that such electronic devices have been traditionally fabricated using technologies such as lithography and printing that generally only work on flat surfaces, 2D to 3D shape sifting could facilitate the incorporation of electronic devices into 3D mechanical metamaterials whose mechanical functionalities originate from their rationally designed 3D micro-architecture.

In addition to multi-functionality, 2D to 3D shapeshifting could also be exploited to create adaptivity in mechanical metamaterials. This concept has been particularly explored in the context of origami-inspired metamaterials where changing the parameters of such origami types as Miura-ori has been used to modify the mechanical behavior of mechanical metamaterials [111, 112].

Depending on whether or not the individual blocks, which move out of their original plane of existence, are deforming themselves, 2D to 3D shape-shifting materials can be categorized as rigid or deformable origami. In what follows, we will discuss both of those categories.

## 4.1. Rigid origami

One of the pathways for 2D to 3D shape transformation is rigid origami where (theoretically) rigid pieces of material, which are joined together by (active) hinges, fold out of the plane to create a 3D object. From the viewpoint of mechanics, this is the simplest form of shapeshifting, because one only needs to solve kinematic equations to describe the folding behavior of rigid origami. Computational approaches that solve the kinematic equations considering the imposed constraints and avoiding collisions during the folding process can be implemented to design origami constructs and devise the folding strategy. Within the context of 4D printing, the problem of creating shapeshifting behavior reduces to that of creating active joints (hinges) that can move out of the plane and carry the rigid parts of the structure with them to create a pre-defined 3D object. The active hinges themselves deform during the folding process, which is why the 4D printing of active hinges themselves requires the application of some of the shapeshifting strategies that we will discuss in the following subsection (*i.e.*, out-of-plane deformations).

Given the kinematic limitations of rigid origami, one fundamental question to ask is: 'what 3D shapes can be realized using rigid origami?' As far as mechanical metamaterials are concerned, some specific geometries are of greater practical importance. For some of those categories, the answer to this fundamental question is immediately clear. For example, there are a large number of origami-inspired mechanical metamaterials [22, 111-115] for which we directly know that the required shapes can be realized using rigid origami approaches. Another important category of mechanical metamaterials employs micro-architectures that are essentially either beam-based or sheet-based lattice structures (or a combination thereof). Whether or not those geometries can be folded from a flat state is more challenging to answer. To answer this question, it is important to consider beam-based and shell-based lattice structures separately.

Beam-based lattice structures are generally not intrinsically curved and are, thus, developable. The Euclidean nature of such structures means that, from the geometry viewpoint, there is no reason that they cannot be folded from a flat state. Indeed, beam-based lattice structures can generally be considered polyhedral with many vertices and edges. In 1999, Demaine presented a mathematical proof for the theorem that every polyhedron can be folded from a sufficiently large piece of paper [116]. The proof, however, did not lead to a practical way of folding polyhedra from a piece of paper (or any flat object). Moreover, the proposed algorithms did not preserve the topology of the 3D shape, as discontinuities (*i.e.*, holes) were introduced during the folding process. A breakthrough occurred when in the early years of the current century Tomohiro Tachi developed the software program "Origamizer" (https://origami.c.u-tokyo.ac.jp/~tachi/software/). This program uses a heuristic algorithm to calculate the crease pattern associated with any 3D shape described by a polyhedral model. The crease patterns resulting from the program are widely used and are generally very practical. Another important development was in 2017 when Demaine and Tachi presented a practical algorithm that not only could fold any polyhedral model from a sufficiently large piece of paper but also preserved the topology of the resulting 3D object [117].

The aforementioned algorithms could, in principle, be applied to beam-based lattice structures. However, they usually do not lead to practical structures for 4D printing. That is because "foldable" is not the same as "self-foldable". Self-folding imposes additional requirements, such as a limit on the magnitude of the required forces and the complexity of the folding sequences. Moreover, folding physical structures with finite thickness is more complex than folding paper, which has a negligible thickness. Finally, to create mechanical metamaterials with sufficient structural integrity, the folded structures need to lock into each other or otherwise provide a load-bearing mechanism. That is why multi-story lattice structures similar to those used in mechanical metamaterials have been rarely folded from a flat state. Recently, we proposed a folding strategy for the self-folding of beam-based multi-story lattice structures [62]. We proposed three different variants of the folding strategy to cover a large number of space-filling polyhedra [62]. Depending on the type of the unit cell on which the lattice structure is based, a different variant of the folding strategy can be used (Figure 4a) [62]. The key elements of the folding strategy that make it applicable to multi-story lattice structures are 1. a limited number of simultaneous folding operations, 2. thickness accommodating features, and 3. incorporated locking mechanisms. The design of the panels and the sequence of the folding operations should be optimized to allow for all features to be included in the resulting lattice structure.

Sheet-based lattice structures tend to be intrinsically curved and are, thus, non-developable. It is, therefore, not possible to fold an initially flat object into sheet-based lattice structures without imposing in-plane deformations, which are not admissible in rigid origami. Even though it is not possible to create intrinsically curved surfaces within the tools of rigid origami, it is possible to fold surfaces that approximate curved surfaces. For many practical purposes, those approximate solutions may be sufficient. The possible folding techniques that can be found in the literature include various types of origami tessellation techniques (*e.g.*, Miura-ori tessellations, Triangular Ron Resch tessellation, and square waterbomb tessellations), tucking molecules, and curved-crease origami [85]. A full review of such techniques has been presented elsewhere [85]. It is important to realize that the abovementioned techniques generally approximate the global intrinsic curvature of a surface. At a local level, however, the paper remains Euclidean. Extending the folding techniques mentioned here to thick-paper origami, which is our primary interest in the case of mechanical metamaterials is not necessarily straightforward.

Lattice structures based on TPMS are one of the most interesting classes of sheet-based lattice structures and have been widely studied in the literature. 4D printing of initially flat TPMS-based lattice structure presents a major challenge given the fact that TPMS belong to the realm of hyperbolic geometry. Recently, a new approach has been proposed for folding TPMS lattice structures from a flat state that combines a network of skeletal components joined together through a set of complex 3D printed joints with pre-stretched sheets that drive the self-folding behavior of the structure towards a state of minimum energy (Figure 4b) [86]. Curvature distributions measured using micro-computed tomography (micro-CT) have shown that this approach results in a good approximation of a minimal surface including near-zero mean curvature measured for most places within the folded construct. Moreover, as opposed to the previously mentioned approaches, a combination of a skeleton and a pre-stretched sheet makes it possible to create local intrinsic curvatures. That is because the pre-stretched sheet undergoes in-plane deformation itself. In that sense, combining a network of rigid skeletons with a pre-stretched sheet is perhaps an intermediate concept between fully rigid origami and out-of-plane deformation discussed in the next sub-section.

While the kinematically valid approaches discussed in this sub-section are all promising for creating the geometries required for the fabrication of mechanical metamaterials, the actual realization of these techniques using 4D printing techniques involves several practical challenges including the generation of large enough folding forces and the incorporation of time delays in the folding sequence to enable 'sequential folding'. The solutions to those challenges are specific to the applied 4D printing approach and the materials under study and are, thus, outside the scope of the current review that focuses on the fundamental aspects of the 4D printing of mechanical metamaterials.

### 4.2. Non-rigid 2D to 3D shapeshifting

2D to 3D shape transformation could also occur through the out-of-plane deformation of a flat object upon the application of a certain stimulus. Even though many combinations of materials, stimuli, and fabrication techniques have been proposed in the literature, the underlying deformation mechanisms usually fall under one of the following categories: bending, buckling, or kirigami (Figure 5-7). While the first two categories of approaches (*i.e.*, bending and buckling, Figure 5) are more abundantly studied in the literature, kirigami-based approaches (Figure 7) are relatively recent. In the cases of bending and buckling, there is a differential response to the stimulus (or a differential application of the stimulus) that leads to the out-of-plane deformation. For the bending approach to work, the differential response should usually be implemented through the thickness (*i.e.*, different through-the-thickness layers respond differently to the applied stimulus) while in the case of buckling, the differential response is usually in-plane (*i.e.*, different areas within the flat construct respond differently to the applied stimulus or are subjected to the different levels of the stimulus). Kirigami-based shapeshifting structures usually work based on (local) instability, which is a different phenomenon as compared to the buckling of the entire or large parts of the flat construct both in terms of the applied stimulus (mostly direct application of mechanical force) and the potentially also the type of instability (*i.e.*, local vs. global instability). Given the fact that in-plane deformation could take place within the flat constructs, it is possible, at least in principle, to transform the shape of the flat construct into both developable and non-developable surfaces. The in-plane deformations are more clearly seen in the case of the buckling approach but could be created in the case of other approaches as well. Since bending, buckling, and kirigami-based approaches all involve (complex) deformations in the underlying materials, solving the kinematic equation is not sufficient to describe the shapeshifting behavior of the 4D printed materials. Indeed, the constitutive equations characterizing the effects of finite deformations, material nonlinearity, strain rate dependency, and permanent deformations should be also solved, together with the equilibrium and kinematic equations.

The bending strategy for shapeshifting requires a through-the-thickness differential response to the applied stimulus (Figure 5a). The simplest case is when the material is made of two layers one of which is (more) responsive to the stimulus and elongates (more) while the other is less responsive and resists the elongation of the first layer. This will result in the out-of-plane deformation of the bi-layer construct, as explained by Timoshenko's bi-metal theory [118, 119]. According to that theory, the shape of the resulting self-bent strip is dependent on several parameters including the mechanical properties of both materials, their thicknesses, and the severity of the differential response of both layers. More layers or a more gradual change of the material behavior through the thickness could be used to gain better control over the shapeshifting behavior of such constructs.

Some other shapes, such as helices (Figure 6a-b), could be also obtained through this approach by introducing anisotropy into the design of the bi-layer constructs such that the responsive layer elongates more in one direction as compared to the other. In the case of such anisotropic responsive materials, changing the angle between the main axis of elongation of the responsive material and the longitudinal axis of the bi-layer strip could be used to create self-twisting behaviors [33, 119]. The geometry of the resulting helix can be adjusted through a combination of the previously mentioned parameters of the Timoshenko's bi-metal theory as well as the angle between both axes [33, 119].

Obtaining more complex shapeshifting behaviors than self-rolling and self-twisting is possible through two different strategies. First, the self-bending and self-twisting shapes described before could be used as the building blocks of more complex shapes in which an array of such basic elements are spatially arranged. Second, the self-bending and self-twisting elements could be combined either with the other strategies mentioned in this section (*i.e.*, buckling or kirigami) where material deformation takes place, or they can be used as the active hinges of rigid origami constructs, which were discussed in the previous sub-section.

The buckling approach works based on in-plane compressive loads that are generated within certain areas of the flat constructs and cause the out-of-plane deformation of some more passive areas within the same flat construct, as they oppose the expansion of the responsive regions and ultimately buckle (Figure 6b). The main difference between this category of shapeshifting materials and those working on the basis of bending is, thus, the in-plane arrangement of the differential response as opposed to the through-the-thickness arrangement seen in the case of bending. Similar to the case of bending, functional gradients could replace the binary choice of responsive *vs.* passive material to improve our control of the shapeshifting behavior (Figure 6b). Moreover, the local application of the stimulus could be used to achieve a similar response as in the case of the in-plane distribution of materials with differential response to the applied stimulus (Figure 6b). Finally, an external piece of responsive material positioned on top of the flat construct could be used to apply the compressive force and induce the buckling of the 2D object (Figure 6b).

The fact that no through-the-thickness gradients are needed to be created in the buckling approach means that 4D printing techniques with relatively low printing resolutions may be acceptable for creating shape-shifting objects. The thickness of the flat construct also plays an important role particularly given the different relationships between the thickness of the flat construct on the one hand and the elastic energy associated with in-plane deformations ($E \propto t$) and out-of-plane bending (buckling) ($E \propto t^3$) on the other [119]. For low thickness values, the bending energy is very small as compared to the energy stored when buckling out of the plane while bending could store much energy for larger thickness values. The balance between the stretching and bending terms determines the amount of buckling *vs.* in-plane deformation and the ratio of the principal curvatures of the flat construct. For thick enough plates, intrinsically curved shapes are both possible and are shown to be feasible to obtain using 4D printing techniques [34]. Various combinations of the shapes of the stimulus-responsive and passive as well as different types of boundary conditions could be used to realize complex shapeshifting behaviors (Figure 6c).

Kirigami-based approaches for shapeshifting behavior are usually based on mechanical instability that occurs either locally or globally. Moreover, the stimulus used in the case of kirigami-based design is primarily the direct application of mechanical forces. The fact that mechanical forces are used in place of the stimulus-responsive behavior of specific materials means that many more materials could be used for the 4D printing of mechanical metamaterials. Moreover, purely mechanical behaviors are usually very scalable, meaning that the translation of the applied design principles, the shapeshifting behavior, and the obtained 3D geometries to microscale and nanoscale devices is less challenging than other techniques for which very fine 4D printing resolutions and very specific materials may be needed to create micro-/nanoscale devices. Finally, the multi-story designs that are used in many mechanical metamaterials cannot be easily folded from a flat construct using only the forces that are generated by the material themselves. The direct application of mechanical forces makes it possible to fold complex multi-story shapes starting from a flat object.

The shape and positioning of the cuts are crucial in the design of kirigami-inspired flat constructs. Depending on whether the type of mechanical instability is global or local, the kirigami-inspired 2D to 3D shape-shifting can be classified into two major categories. In the case of global buckling, the geometry of the cuts and their positioning are designed such that the entire flat construct or large parts of it experience buckling upon the application of compressive forces (*e.g.*, see [120, 121]). The design of such structures, thus, requires the (computational) prediction of the nonlinear post-buckling behavior of the flat structure. On the other hand, kirigami-inspired structures that work based on local buckling exploit the local buckling of small regions within the flat construct to create out-of-plane deformations. The advantage of such an approach is its modularity, as multiple modular elements whose response is individually known could be connected to create a complex shape-shifting behavior and fabricate the kind of multi-story lattice structures that are needed for the realization of mechanical metamaterials (Figure 7) [68]. The application of complex computational models for the prediction of the post-buckling behavior of the constructs may not be required as the behavior of each module is predictable and remains unchanged regardless of how the various moduli are organized within the flat construct. Finally, the kirigami-inspired structures that work based on local buckling could be activated using globally applied 'tensile' forces (*e.g.*, see [68]) instead of the compressive forces that are usually required for global buckling. The application of tensile forces can usually be more reliably controlled, as there is no competition between the global and local buckling modes of the material that could switch the type of buckling depending on the severity and type of imperfections present in the structure.

## 5. 3D TO 3D SHAPE TRANSFORMATION

The 4D printing of objects for 3D to 3D shapeshifting has received relatively less attention than 2D to 3D shapeshifting. The main advantage of a 2D object, namely the possibility to use technologies that primarily work on flat surfaces to incorporate multi-functionality and adaptability, is not relevant for 3D printed objects whose shape is 3D to start with. There are, however, other good reasons why one may be interested in 3D to 3D shapeshifting. Some examples of the applications for which 3D to 3D shape transformation is relevant include actuators, deployable structures, and shape morphing objects. Some of those applications impose certain requirements regarding the shapeshifting behavior of the 4D printed materials. For example, actuation imposes specific requirements regarding not only the speed and reversibility of the shape transformation process but also the forces generated during that process and the fatigue behavior of the parts of the 4D printed object involved in shape transformation. Similar to the classification used in the previous section, we will divide our discussion of 3D to 3D shapeshifting into 1. "rigid shape morphing" where the different parts of the 4D printed object do not undergo substantial deformation but rather move with respect to each other, and 2. deformable-body shape morphing where the objects substantially deform during the shapeshifting process. In the two following subsections, we will discuss each of those categories separately.

## 5.1. Rigid shape morphing

As opposed to 2D-3D shape-shifting where the mathematical problem of shapeshifting is well-defined and has been extensively studied, 3D-3D shapeshifting is less clearly defined and studied. In 2D to 3D shape-shifting, questions like 'which 3D shapes can be folded from a flat state' and 'what 3D surfaces are developable' affords the field of study with a certain degree of order and classifies the different approaches along the lines of the answers to those fundamental questions of foldability. In 3D to 3D shape-shifting, there are hardly any equivalent research questions. Instead, it may be useful to categorize the 3D to 3D shape-shifting approaches in two main categories. In the first category, the start and final shapes of the architected material are limited to a few pre-determined possible shapes. This is, for example, the case in multi-stable and deployable structures. Multi-stable mechanisms and structures [114, 122-125] are the prime examples of such an approach. There are usually a few (*e.g.*, two in the case of bi-stable structures) stable states of the 3D structure (with their associated shapes and mechanical properties) between which the shape-shifting occurs. The switching from one stable state to another is often associated with mechanical instabilities, such as snap-through instability. Architected materials based on (snap-through) instability and related concepts [1, 126-131] exhibit a rich set of mechanical behaviors, including adaptive stiffness, negative stiffness, and a negative Poisson's ratio. There is, therefore, a natural connection between the design concepts that are used for affording the material with unusual mechanical properties and the design concepts used to create the shapeshifting behavior. The other important examples of rigid shapeshifting are deployable structures that are used for various applications, including space structures [132-135] and orthopedic meta-implants [136, 137]. Deployable structures are compact before their deployment. The compactness of such structures means that they are easier to transport (in the case of space structures) or are minimally invasive (in the case of orthopedic meta-implants [137]). Some deployable structures use multi-stability as their main working mechanism. However, other working principles can be used too. The 3D-3D shape-shifting materials whose start and end shapes are limited in number and are pre-determined offer the advantage of predictability and a more streamlined design process. The shapeshifting behavior can be designed by solving the kinematic equation of the movements of the individual elements that make up the structure. Even when material deformations are involved, for example, in the mechanisms working based on snap-through instability, the deformations are generally localized and predictable and could be implicitly incorporated into the kinematic equations. Consequently, the simulation of the shapeshifting behavior of such structures usually only requires solving the kinematic equations of the motion of their various constituting elements. The limited number of stable states in the case of multi-stable structures and the clear start and end shapes in the case of deployable structures mean that the activation mechanism for shapeshifting could be very simple. For example, it has been shown that an instrument as simple as an inflating balloon could successfully deploy a complex meta-implant working on the basis of bi-stable structural elements (Figure 8a-c) [136].

Another category of shapeshifting structures has been recently introduced that behave similarly to "clay" (Figure 9) [138]. The analogy between these materials and clay starts from the fact that, due to the presence of many joints in the structure of such clay-like materials, they can change their shape more or less arbitrarily. There is, therefore, no clear 'end shape' and arguably also no clear 'start shape'. The structure can then be molded using either external boundaries that define the exterior shape of the object or by 'expert hands'. Once the desired shape is achieved, a locking mechanism is activated to freeze the shape of the object by limiting the degrees-of-freedom of the structure. This locking mechanism is, therefore, an important concept without which the clay-like structures will be of no or limited use. Both the 'joints' constituting the structure and the 'locking mechanism' can be categorized as being either 'kinematic' or 'complaint' (Figure 9). Whereas the elements joined through kinematic joints move with respect to each other, those connected using complaint joints change their position due to the deformation taking place in their connecting links. Similarly, a kinematic locking mechanism constrains the movement of the different joints using kinematic obstacles while a complaint locking mechanism uses the elastic forces resulting from the deformation of some deformable bodies to constrain the movement of (some of) the joints. Even though there are some deformations taking place in such kind of clay-like materials, they are generally very local and predictable, which is why the most important governing equations are still kinematics in nature and that these clay-like materials are categorized under 'rigid shape morphing'.

Regardless of whether or not the start and end shapes of the shape-morphing object are limited in number and are predictable, the challenges associated with the 4D printing of such structures are more or less the same. In particular, the 3D printing of moving joints, multi-stable elements, and locking mechanisms is a major challenge. That is because of the small features and clearances that need to be printed at a small scale, at a large number, and without the help of support structures. The requirement 'support-free printing' is the most difficult one to meet particularly when using metals [138]. As long as polymeric materials are concerned, support-free printing is relatively straightforward to realize, because it is possible to print the structure using multiple materials of which one (*i.e.*, support material) can be dissolved or otherwise removed. That is not the case for metals, where high-resolution multi-material printing remains to be challenging and prohibitively difficult outside some isolated experiments (*e.g.*, [139-142]). The options for the multi-material printing of metallic materials further decreases when requiring that one of the materials should be dissolvable while the other material should not be affected by the dissolution process. Indeed, no or limited cases of such printing techniques have been demonstrated at the level required for the fabrication of miniaturized joints and locking mechanisms. On the other hand, the use of metallic materials is often required for some categories of rigid shape-shifting materials particularly when, similar to the case of mechanical metamaterials, the mechanical response of the printed materials is of primary importance. Under such circumstances, the use of metallic materials is often the only choice to ensure that the resulting materials exhibit enough stress-worthiness to be of utility in practical applications. That is partially because creating deployable and clay-like structures usually necessitates the use of as little material as possible to accommodate the movements of the structural elements of the structure with respect to one another. Given the minimum amount of material available for carrying the mechanical loads, it is important to fabricate the structures from materials with high mechanical properties (*e.g.*, metals) and to use all the applied material as efficiently as possible. There are, therefore, additional design considerations that need to be taken into account when developing mechanical metamaterials with rigid shape morphing behavior. As a consequence of such additional design considerations, it may be necessary to re-design the common mechanical elements, such as kinematic joints to make them support-free 3D printable and to increase their stress-worthiness. These are topics that have recently started to attract the researchers' attention and remain largely unexplored.

## 5.2. Deformable-body shape morphing

As long as the overall shape of an object (and not the detailed micro-architecture of an architected material) is concerned, deforming a 3D shape to most other 3D shapes is possible. The main difficulty lies in activating and guiding the shapeshifting process in such a way that the applied deformations or forces lead to the desired changes in the shape. As previously mentioned, the 3D to 3D shape-shifting behavior of mechanical metamaterials is primarily useful for the design of actuators, soft robots, medical devices, and (embedded) mechatronic systems. In most of such applications, the induced shape transformation needs to be reversible. In the context of 'deformable-body' shape morphing that usually means that the underlying materials should be able to undergo large elastic deformations repeatedly. Elastomers and other rubber-like materials are particularly suited for such applications.

Guided shape transformation through elastic deformations can be realized using various strategies. The key to all those approaches is to devise strategies for guiding the elastic forces such that they create highly predictable and precisely controllable shape transformations. One obvious strategy is to create (rigid) boundary conditions that locally constrain the elastic deformations and guide the overall shape-shifting behavior of the mechanical metamaterial. While this approach can be relatively easily applied to some simpler designs where access to all the relevant points of the structure is relatively easy to realize, it is very challenging to apply to fully 3D lattice structures that are often used for the design of mechanical metamaterials. That is because the internal areas within the lattice structure, which play important roles in determining the shape transformations resulting from a set of applied elastic forces, cannot be easily accessed to apply/modify the internal boundary conditions.

Another approach that can be used either independently or in combination with the application of boundary conditions is to use the buckling modes of soft lattice structures to achieve a specific shape-shifting behavior (Figure 10). The advantage of this approach is that complex shape transformations can be realized without any need for complex and locally distributed boundary conditions. Moreover, the buckling load and buckling shape of lattice structures are highly predictable and can be adjusted through both geometrical design and spatial distribution of material properties [9, 143-147]. Several studies have used this type of mechanical instabilities to create mechanical metamaterials that exhibit specific types of shape transformations under compression. Some variants of this approach enable creating shape transformations also under tension [68, 130]. In both cases, the stimulus that triggers shape transformation is the direct or indirect application of mechanical forces. The disadvantage that many such approaches face is that it is not possible to go beyond the first mode of buckling of the lattice structures, as the first buckling mode occurs first and prevents the other modes from playing a role in the shape transformation. This limits the envelope of the achievable shapes because the higher buckling modes are usually much richer than only the first buckling mode. It is, therefore, desirable to activate (a combination of) the higher buckling modes of such lattice structures. Recently, some studies have proposed some relatively simple but highly effective strategies for that. In particular, the geometry of the lattice structures can be modified by adding small purposefully introduced imperfections that pre-dispose the lattice structure to activate its higher modes of buckling instead of the first buckling mode (Figure 10a) [144]. Very complex shapes combining several buckling modes of soft mechanical metamaterials can be achieved using this approach. The advantage of this approach is that no complex boundary conditions are required.

Finally, a rational approach to the spatial distribution of the mechanical properties within a lattice structure could be used to program the shapeshifting behavior of mechanical metamaterials. In particular, the spatial distribution of the Poisson's ratio lends itself well to the programming of shapeshifting in architected materials [148]. The advantage of this approach is its simplicity and predictability. However, it brings some practical design challenges as unit cells with highly different designs and dimensions may need to be combined to achieve the desired distribution of the Poisson's ratio. Since such unit cell designs are not necessarily compatible with each other, additional structural elements may need to be added to bridge one type of unit cell with another. The addition of such additional structural elements is not necessarily straightforward to implement in an automated manner and could also lead to changes in the mechanical behavior and, thus, shapeshifting behavior of the lattice structure. As a result, sudden changes in the shape (*i.e.*, sharp angles and curves) may not be accurately captured using this approach [148].

In addition to the approaches discussed here, the methods discussed in the previous sections could be also used for 3D-3D shape transformations. Some of the examples include those based on kirigami designs and the differential response to the applied stimuli (*e.g.*, high temperatures or swelling in hydrogels). Finally, we did not make a distinction between 3D-3D and 2D-2D transformations, as they are often not fundamentally different in terms of the underlying concepts.

## 6. DISCUSSIONS AND FUTURE RESEARCH DIRECTIONS

4D printing of mechanical metamaterials involves challenges on multiple fronts that need to be addressed before mechanical metamaterials with relevant properties and functionalities can be realized using this fabrication technique. The most important challenges facing researchers is different depending on whether 2D to 3D or 3D to 3D shape transformation is concerned. In the case of 2D to 3D shape transformation, the three main challenges include fabricating lattice structures with 1. enough design freedom (*i.e.*, form-freedom challenge), 2. a large enough stack of unit cells (*i.e.*, multi-story constructs challenge), and 3. small enough size of unit cells (*i.e.*, micro-/nanoscale fabrication challenge). Since 2D to 3D shapeshifting is in most cases used as a way to fabricate multi-functional and adaptive metamaterials, the reversibility of the shape transformation and the material fatigue behavior are of lesser importance. In comparison, 3D to 3D shape transformations, particularly when they are used for actuation and the other types of dynamic applications, often require reversibility and fatigue-resistant materials that can repeatedly undergo shape transformation while marinating their structural integrity and stress-worthiness.

As far as 2D to 3D shape sifting is concerned, our presented review of the literature shows that significant progress has been made regarding the first challenge, namely form-freedom. Several generality results and theorems, as well as practical folding strategies discussed in the previous section of the paper clearly, shows that we have now established that it is possible, at least in theory, to fold the most relevant 3D shapes from a flat structure as long as we allow for some approximations of the final shape. In addition to this generality result that guarantees the existence of a folding strategy but does not necessarily provide a practical one, important developments have been made during the last decade regarding practical folding strategies that could be used to fold complex 3D shapes from a flat state. The progress regarding the other challenges has, however, been relatively limited. Despite the appearance of a growing list of materials, processes, and fabrication techniques in the literature that are proposed to create the desired shapeshifting behaviors, most proposed solutions are not capable of creating multi-story lattice structures with a small size of unit cells. The bulk of current research activities focuses on finding new materials and processes for 4D printing. However, the key to addressing the two remaining challenges may be in changing the perspective and using completely different approaches. The recently proposed approach of using mechanical forces for the folding of miniaturized multi-story lattice structures is a promising candidate that provides novel solutions for the last two challenges faced in 2D to 3D shape-shifting of mechanical metamaterials. It is, therefore, suggested that future research should be conducted to further explore the opportunities offered by mechanical routes to shapeshifting.

As for 3D to 3D shape transformation, reversibility and fatigue-resistance remain the most important challenges that need to be solved. While several approaches have been developed in recent years to enabled untethered, reversible shape-shifting [149], this aspect remains largely unaddressed particularly given the fact that a large proportion of the currently available 4D printing processes only enable one or limited cycles of shape transformation. Even when multiple cycles of shape transformation are possible, it is often the case that the reversibility is only partial given the existing plasticity that leads to permanent, irreversible deformations. Future research should, therefore, address the reversibility of the shape shifting behaviour and develop strategies that allow for a large number of fully reversiblet shape transformations. Once more, mechanical forces could play an important role in achieving such kind of behaviours, as they are more easily controlled through the different length scales and are less influenced by such factors as impurities and defects in the chemical composition and the crystalline structure of materials that may lead to imperfections in the reversible shape shifting behavior of the resulting 4D printed objects. Fatigue-resistance in the shape shifting behavior is another topic that is almost entirely neglected in the existing literature. In order for 4D printed actuators to be of practical utility, they need to be able to undergo tens to hundreds of thousands of shape transformation cycles. However, there is barely any data in the literature as to how many shape transformation cycles could 4D printed objects undergo before losing their stress-worthiness and, eventually, structural integrity. The study of the fatigue behaviour of 4D printed shape shifting objects could, therefore, constitute a fertile ground for future research.

## ACKNOWLEDGEMENTS

This research has received funding from the European Research Council under the ERC grant agreement n° [677575].

**Figure captions**

**Figure 1.** A schematic drawing of the different variants of (a) geometrical and (b) topological 4D printing where either the geometry (subfigure a) or both the geometry and topology (subfigure b) change after the application of a stimulus.

**Figure 2.** (a) The definitions of the principal curvatures, mean curvature, and Gaussian curvature [85]. (b) The various types of geometry depending on the value of the Gaussian curvature.

**Figure 3.** (a) The distribution of the Gaussian curvature within some representative geometrical shapes. (b) The relationship between the distribution of the Gaussian curvature and the genus values of various topological objects.

**Figure 4.** (a) The folding strategies for folding three categories of lattice structures. Depending on the type of the repetitive unit cell used as the basis for the design of the lattice structure, one of these three folding strategies could be used for folding it from a flat state [62]. (b) The tiling strategy that is used for folding hyperbolic geometries, such as TPMS from a flat state [86].

**Figure 5.** The different strategies used for creating out-of-plane deformation using the bending (a) and buckling (b) shapeshifting mechanisms. The green and blue colors refer to materials with different stimulus-response and/or mechanical properties [119].

**Figure 6.** More complex shapes, such as helices could be obtained from a flat state by combining two materials with different anisotropic shrinking behaviors and with different in-plane angles with respect to their axes of anisotropy (a) as well as by combining two sheets of the same materials with different in-plane angles of orientation with respect to their axis of anisotropy (b) [119]. The dimensions of the resulting helices are dependent on multiple design parameters (b) [119]. Similarly, many other shapes can be obtained by rationally combining materials with differential stimulus-response behaviors [119] (c). From top left to the bottom

right, the examples of the shape-shifting objects are adopted from references [150], [151], [152], [153], [154], and [46].

**Figure 7.** Kirigami-inspired approaches for the out-of-plane shape-shifting of flat constructs. (a) The kirigami-inspired contracts could be made either to function as a bi-stale mechanism (green) or undergo local buckling, leading to local out-of-plane deformation (blue) [68]. (b) Combining the kirigami-inspired designs with an elastic material leads to a global out-of-plane deformation that could be harnessed to create 2D to 3D shape-shifting [68]. (c) Using the abovementioned approach, miniaturized multi-story lattice structures could be fabricated [68]. (d) Moreover, it is possible to incorporate flexible electronics into the resulting mechanical metamaterials [68].

**Figure 8.** Multi-layer Russian doll deployable meta-implants working based on multi-stability [136] (a-c). An inflating balloon deploys the implants. The presence of multiple multi-stale layers affords the implants with significant stress-worthiness [136] (d). Such an implant could, for example, be used for minimally invasive treatment of fractured vertebra [136] (e).

**Figure 9.** The concept of metallic-clay is analogous to that of actual clay with shape-morphing and shape-locking stages [138] (a). The shape-morphing behavior of metallic clay is highly dependent on the type and arrangement of the many joints that provide it with numerous degrees of freedom [138] (b). There is, however, a need for either complaint (c) or kinematic (d) locking mechanisms to freeze the shape of the metallic clay structures [138].

**Figure 10.** Introducing imperfections calculated from a post-buckling analysis to pre-dispose the lattice structures such that their higher modes of buckling are activated [144] (a). Various combinations of the different modes of buckling could be also used to create even more complex shape transformations in soft mechanical metamaterials [144] (b). The rational distribution of mechanical metamaterials with different values of the Poisson's ratio enables the approximation of arbitrary shapes [148] (c-d).

**Figure 1**

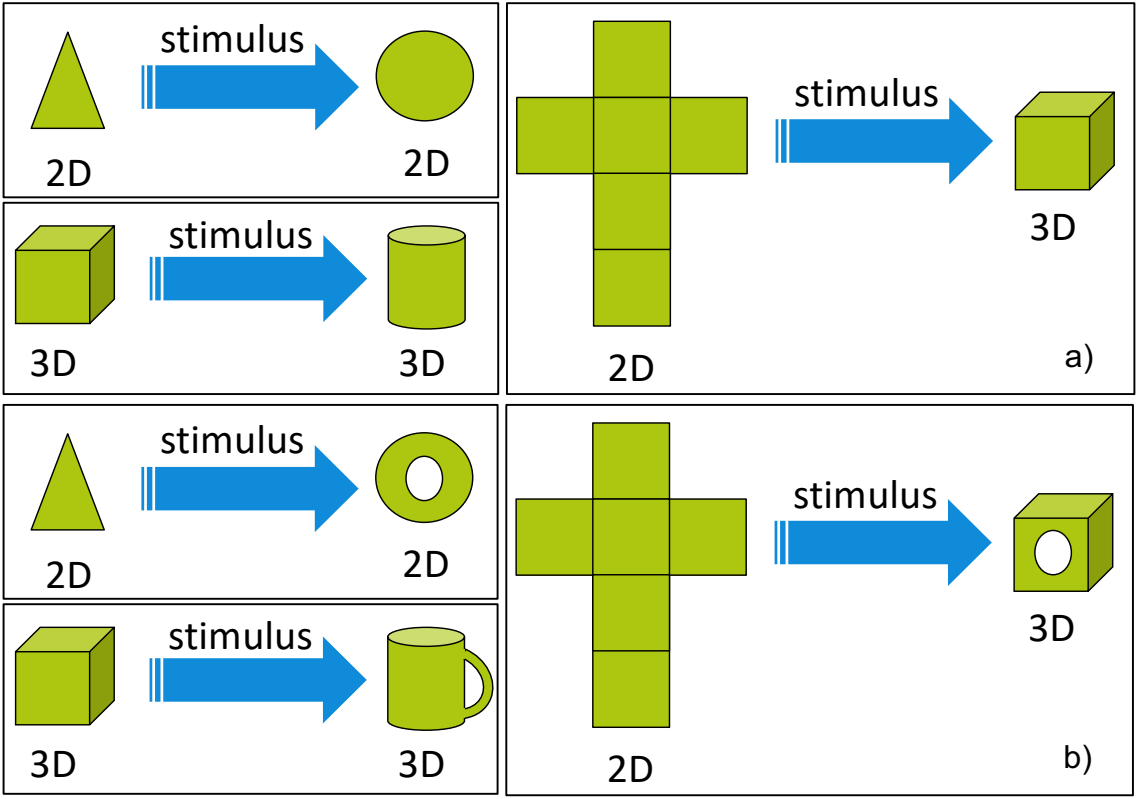

**Figure 2**

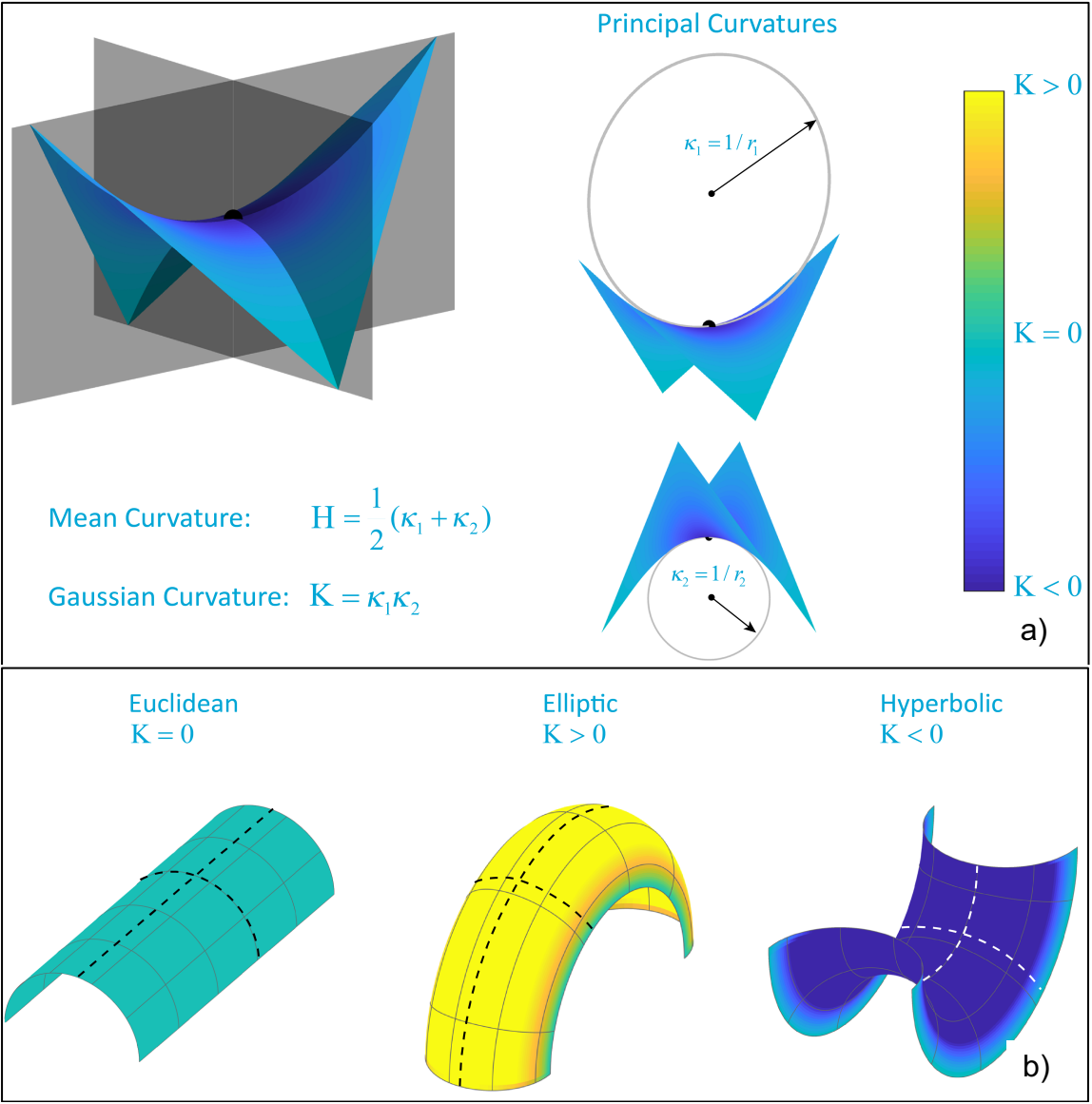

a)

**Principal Curvatures**

$\kappa_1 = 1/r_1$

Mean Curvature: $\mathrm{H} = \dfrac{1}{2}(\kappa_1 + \kappa_2)$

Gaussian Curvature: $\mathrm{K} = \kappa_1 \kappa_2$

$\kappa_2 = 1/r_2$

K > 0

K = 0

K < 0

b)

Euclidean
K = 0

Elliptic
K > 0

Hyperbolic
K < 0

**Figure 3**

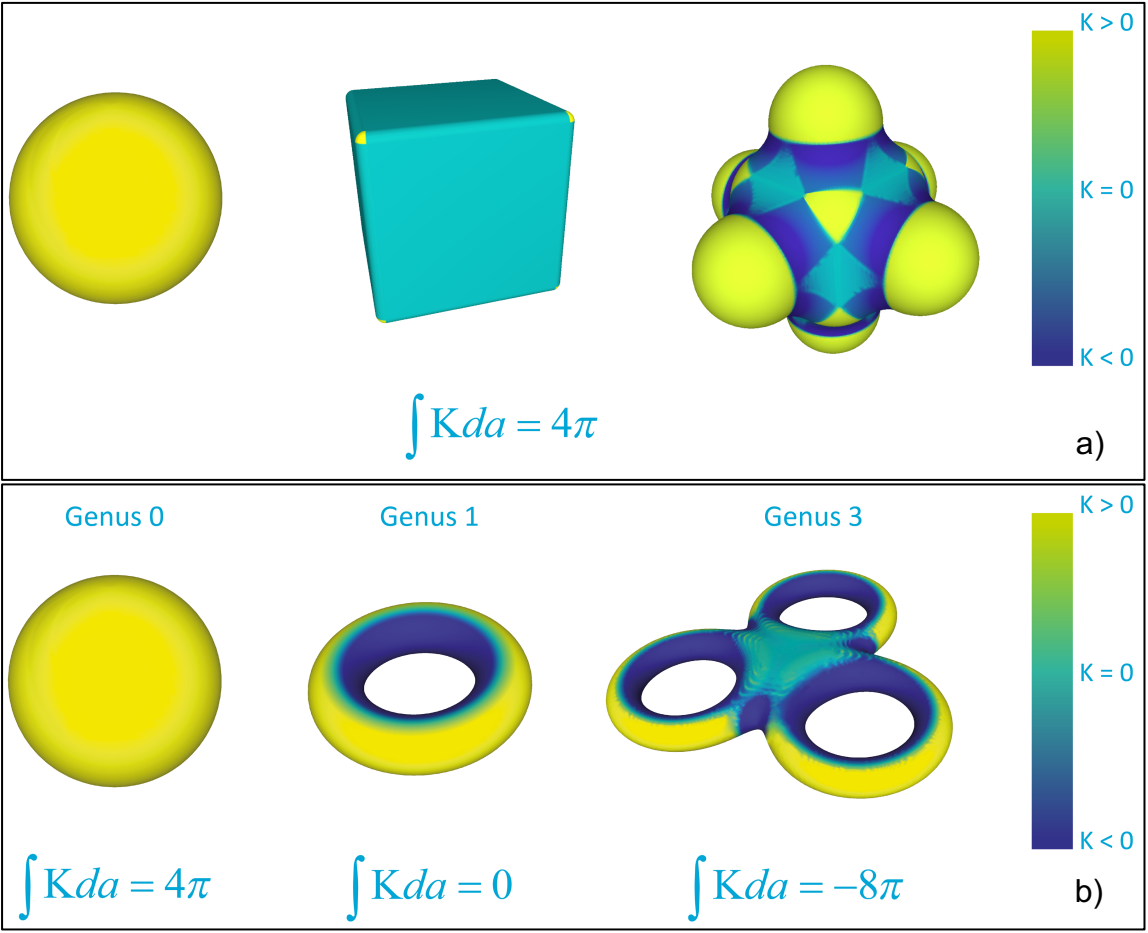

**Figure 4**

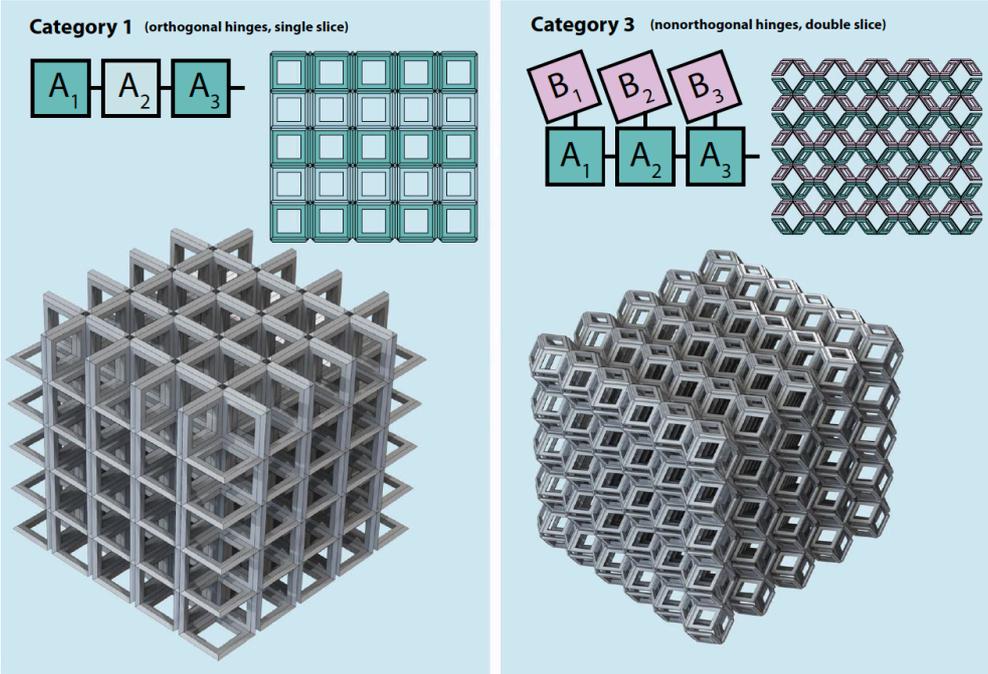

**Category 1** (orthogonal hinges, single slice)

$A_1$ $A_2$ $A_3$

**Category 3** (nonorthogonal hinges, double slice)

$B_1$ $B_2$ $B_3$
$A_1$ $A_2$ $A_3$

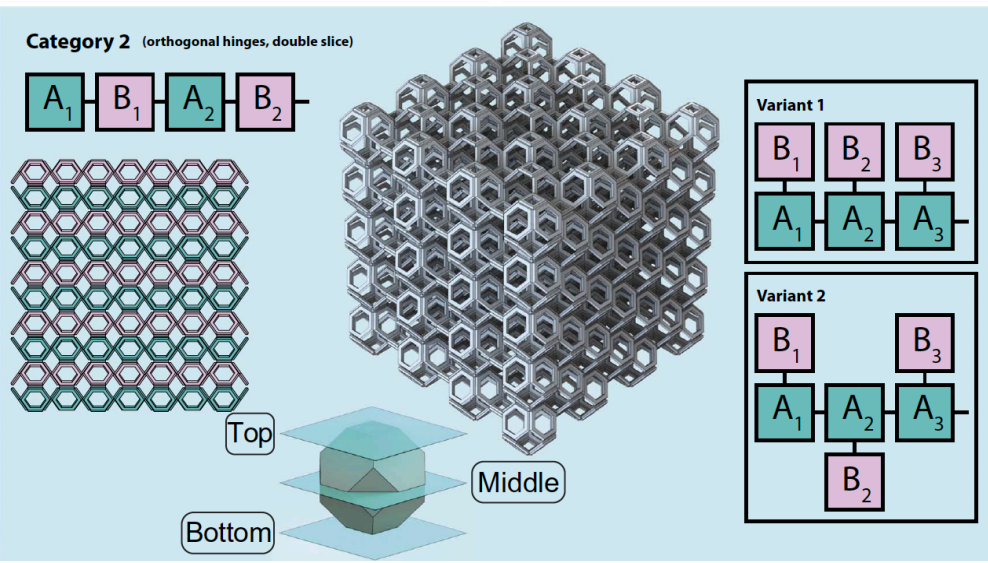

**Category 2** (orthogonal hinges, double slice)

$A_1$ $B_1$ $A_2$ $B_2$

**Variant 1**

$B_1$ $B_2$ $B_3$
$A_1$ $A_2$ $A_3$

**Variant 2**

$B_1$ $B_3$
$A_1$ $A_2$ $A_3$
$B_2$

Top
Middle
Bottom

a)

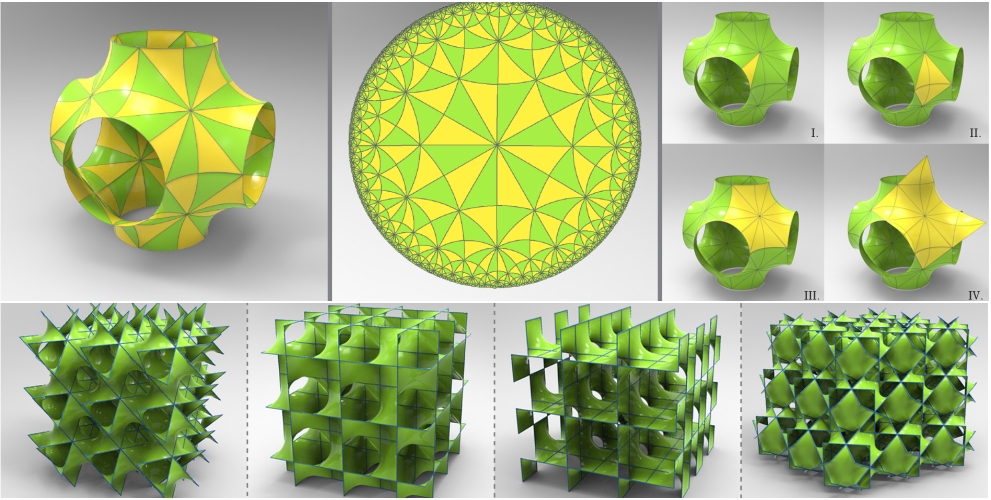

I.   II.
III.  IV.

b)

**Figure 5**

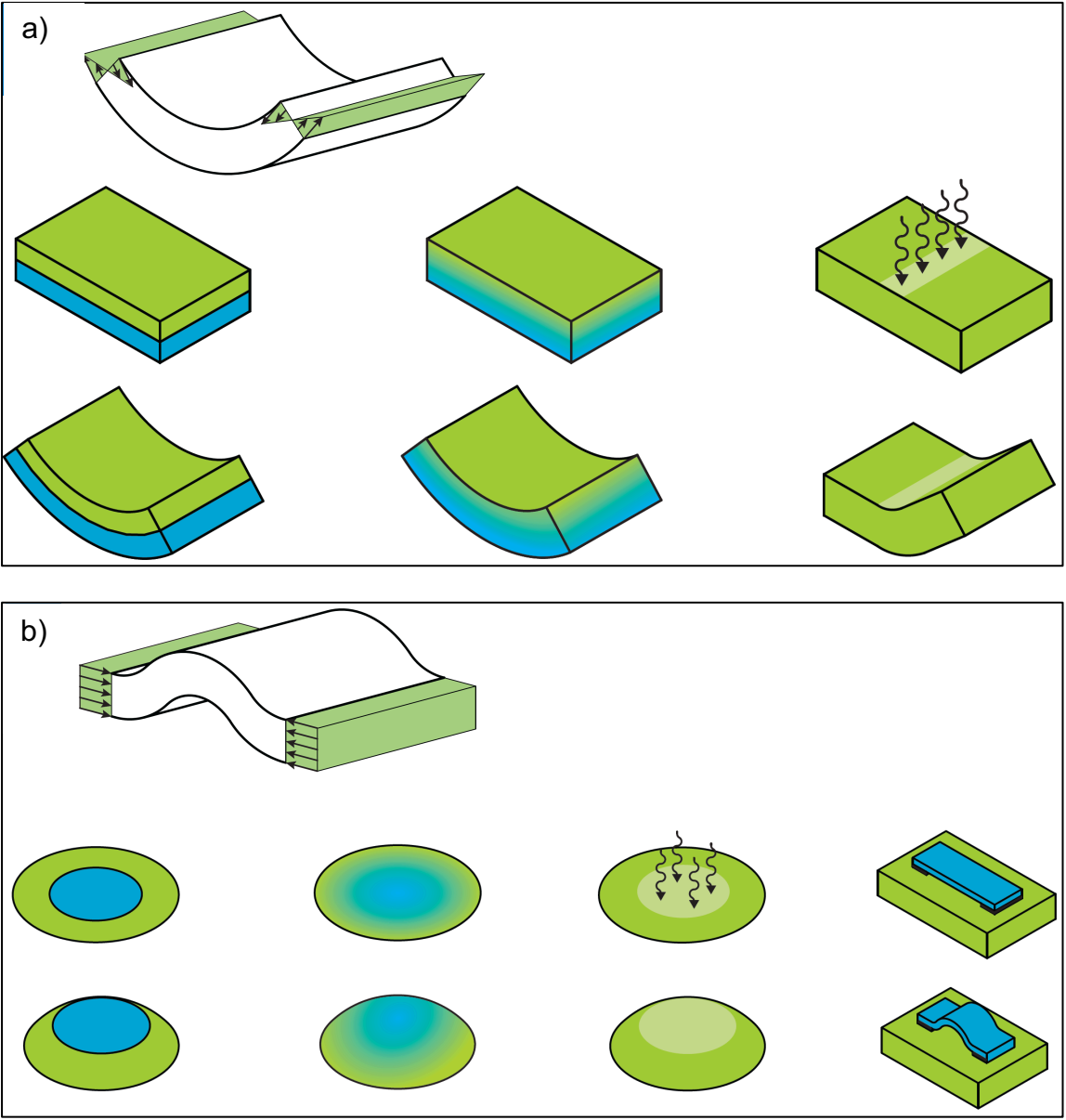

**Figure 6**

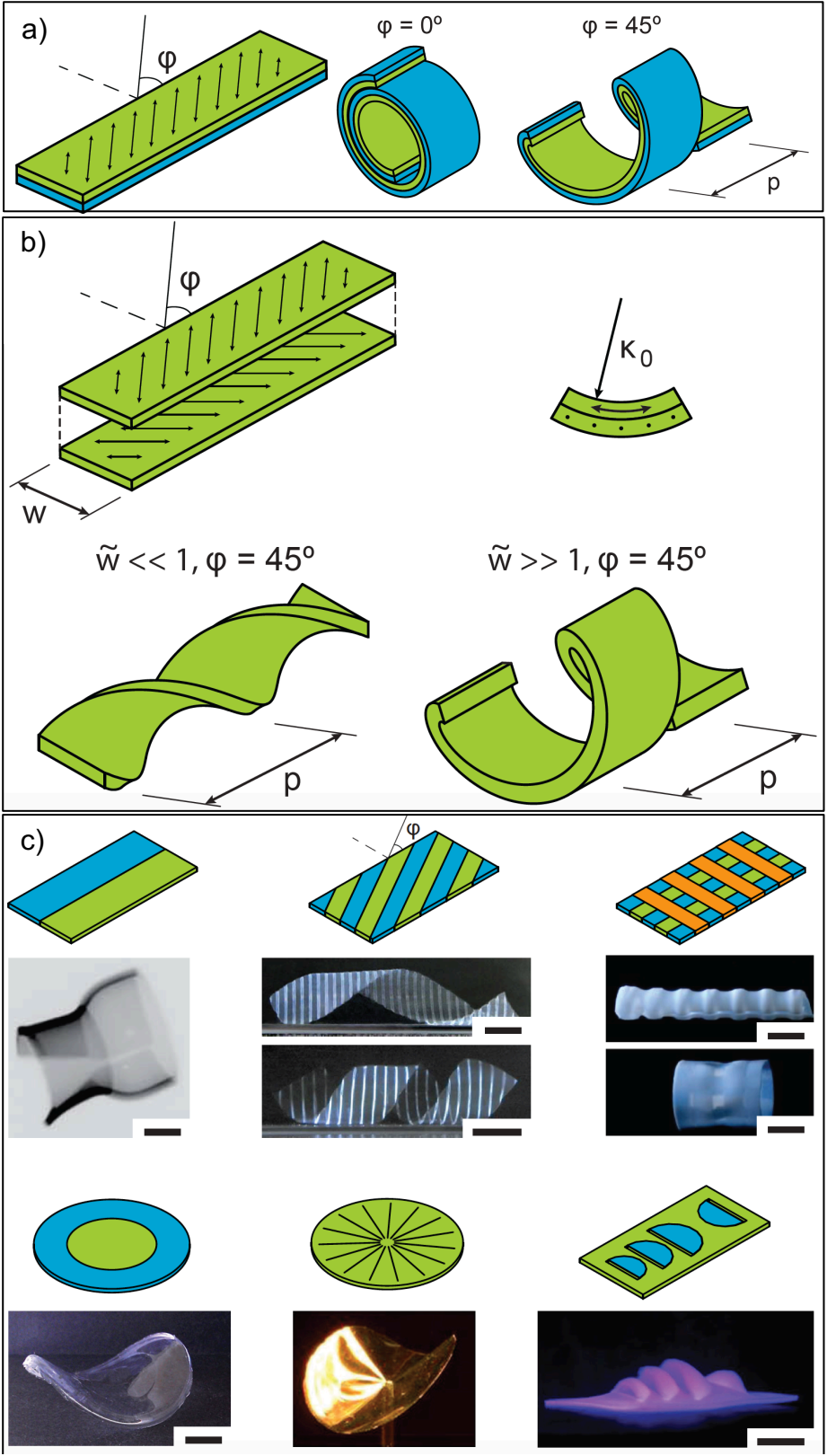

**Figure 7**

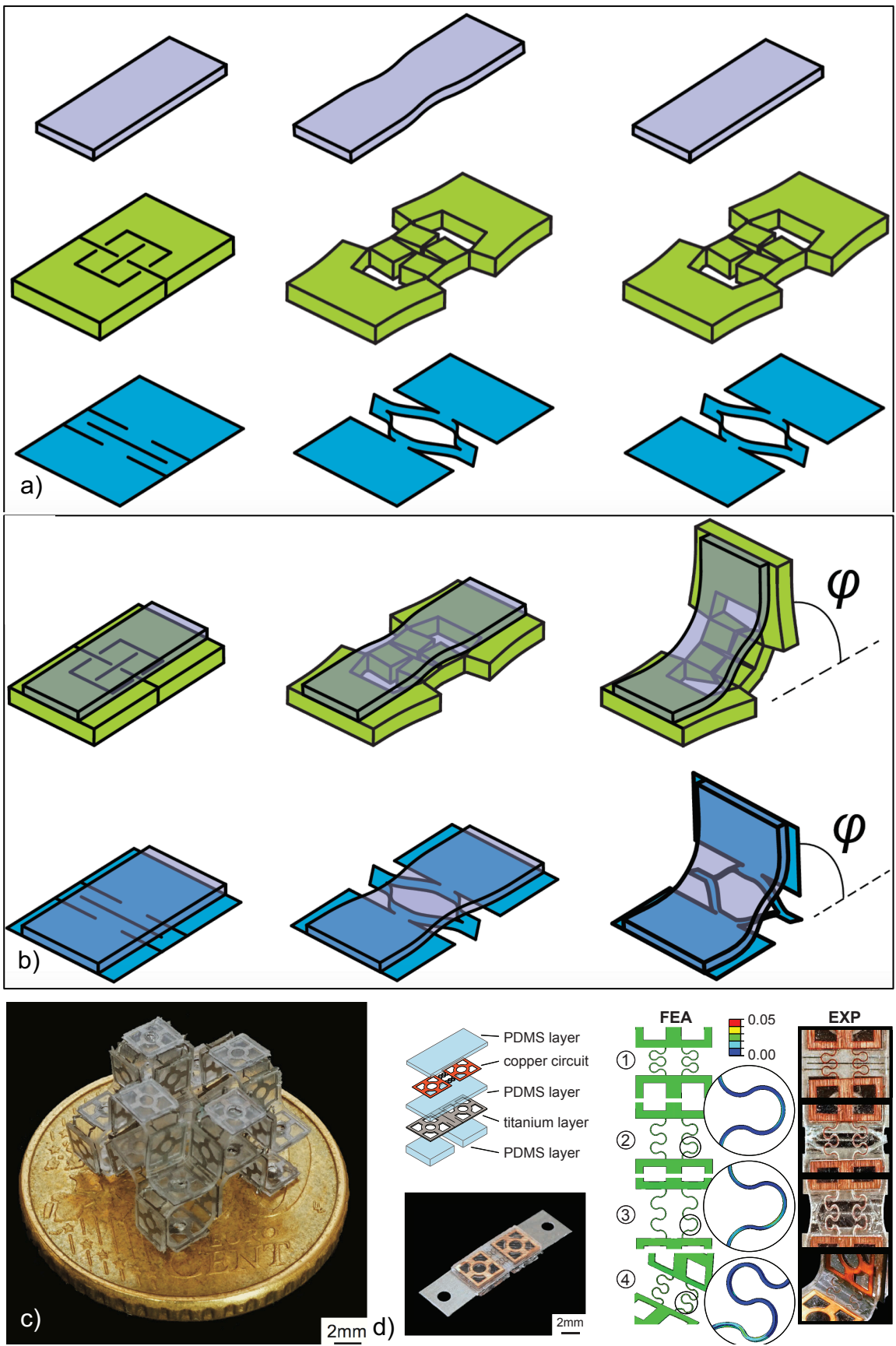

**Figure 8**

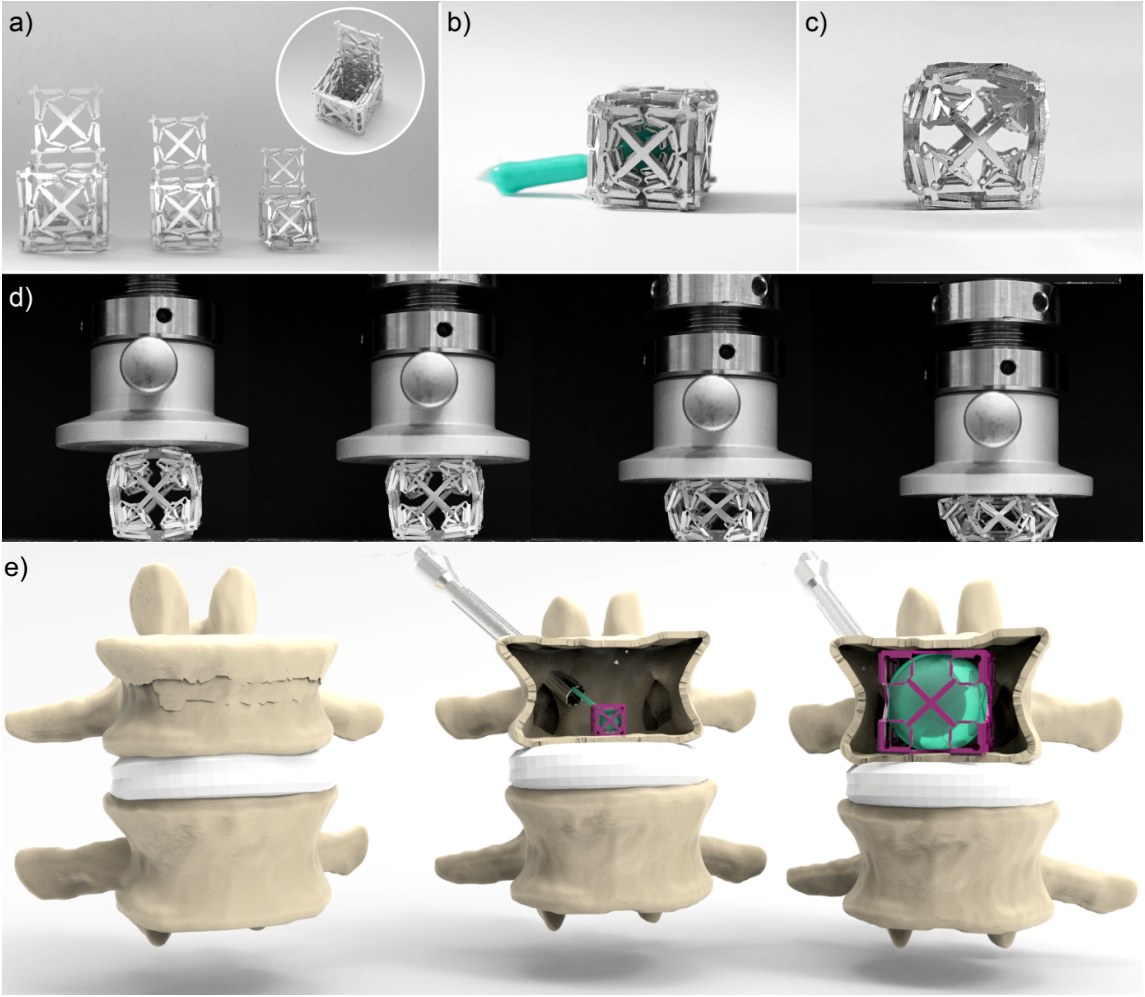

**Figure 9**

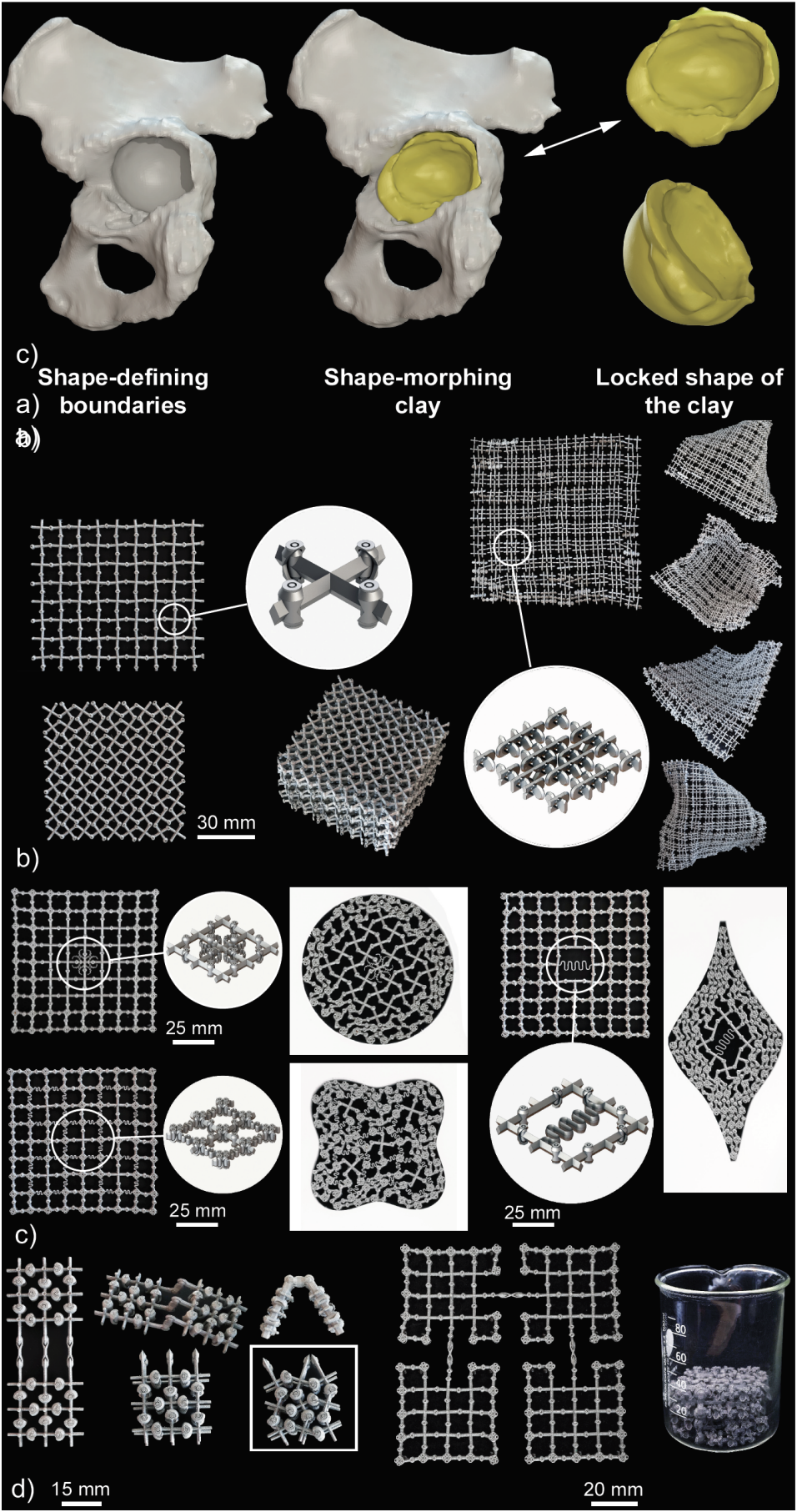

c)
a) Shape-defining boundaries    Shape-morphing clay    Locked shape of the clay

a)

b)

c)

30 mm

25 mm

25 mm

25 mm

d) 15 mm    20 mm

Figure 10

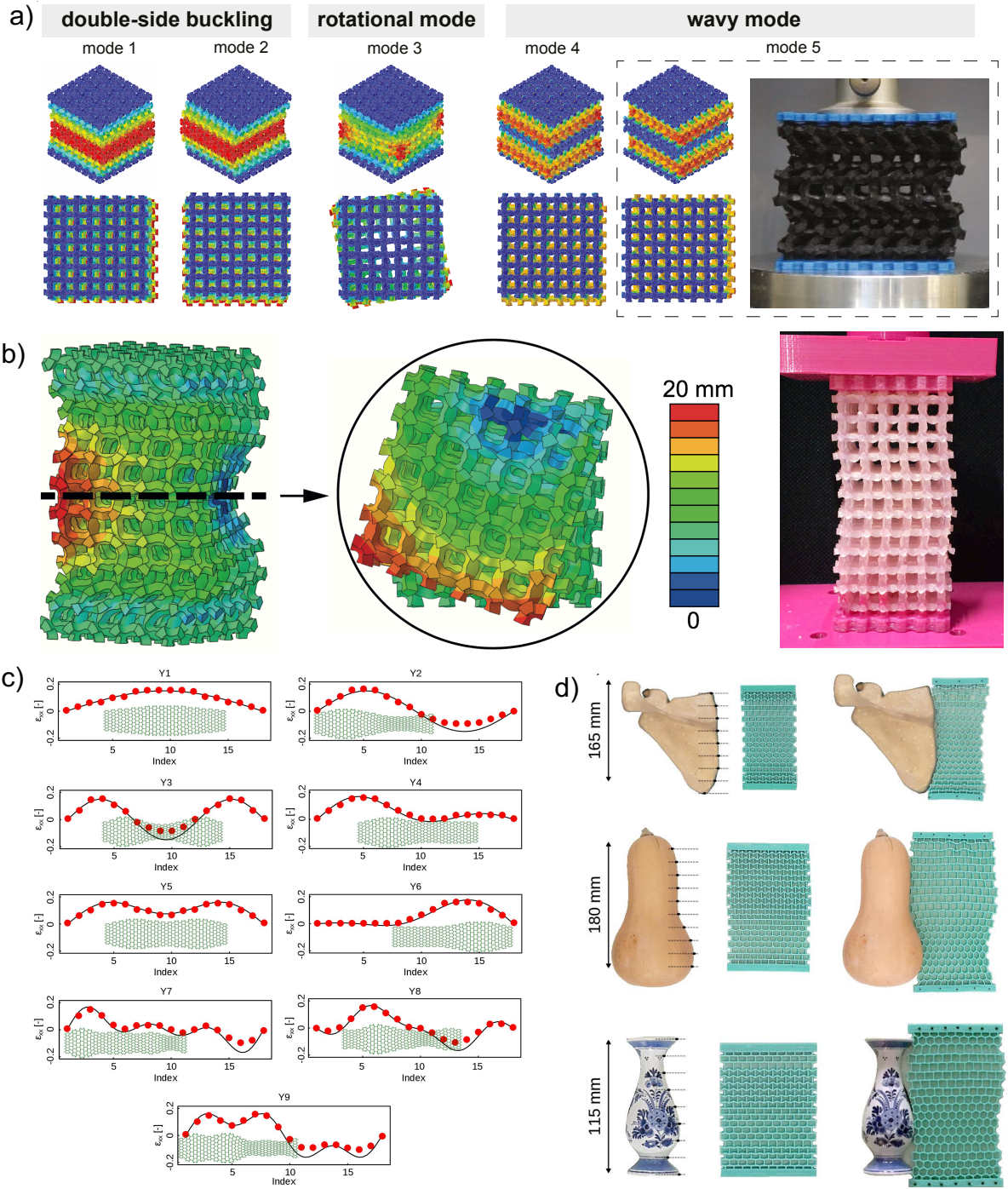